\PassOptionsToPackage{table}{xcolor}
\documentclass[runningheads]{llncs}

\usepackage{eccv}

\usepackage{eccvabbrv}

\usepackage{listings}

\usepackage{wrapfig}
\usepackage{algorithm}
\usepackage{algpseudocode} 

\usepackage{siunitx}

\usepackage{amsmath,amsfonts,bm}
\usepackage{latexsym}
\usepackage{amsmath}
\usepackage{amssymb}
\usepackage{xurl}
\usepackage{bm}
\usepackage{mathrsfs}
\usepackage{mathtools} 
\usepackage{booktabs}
\usepackage{multirow}

\def\eqref#1{equation~\ref{#1}}

\def\1{\bm{1}}

\DeclareMathAlphabet{\mathsfit}{\encodingdefault}{\sfdefault}{m}{sl}
\SetMathAlphabet{\mathsfit}{bold}{\encodingdefault}{\sfdefault}{bx}{n}

\newcommand{\xb}{\mathbf{x}}
\newcommand{\yb}{\mathbf{y}}

\newcommand{\Ab}{\mathbf{A}}

\newcommand{\Db}{\mathbf{D}}

\newcommand{\Gb}{\mathbf{G}}

\newcommand{\Ib}{\mathbf{I}}

\newcommand{\Wb}{\mathbf{W}}
\newcommand{\Xb}{\mathbf{X}}

\ifx\BlackBox\undefined
\newcommand{\BlackBox}{\rule{1.5ex}{1.5ex}}  
\fi

\ifx\QED\undefined
\def\QED{~\rule[-1pt]{5pt}{5pt}\par\medskip}
\fi

\ifx\qed\undefined
\def\qed{~\rule[-1pt]{5pt}{5pt}\par\medskip}
\fi

\ifx\proof\undefined
\newenvironment{proof}{\par\noindent{\bf Proof\ }}{\hfill\BlackBox\\[2mm]}
\fi

\renewcommand{\url}[1]{{\sffamily #1}}

\usepackage{graphicx}
\usepackage{booktabs}

\newcommand{\method}{\textsc{MoECodec}\xspace}

\usepackage[pagebackref,breaklinks,colorlinks,citecolor=eccvblue]{hyperref}

\usepackage{orcidlink}

\usepackage{booktabs}
\usepackage{xcolor}
\usepackage{multirow}
\usepackage{pifont}

\usepackage{adjustbox}

\begin{document}

\title{\method: Image Compression for joint human and machine perception via Mixture-of-Experts} 

\titlerunning{\method}

\author{Jiancheng Zhao\inst{1} \and
Xiang Ji\inst{1} \and
Yifan Zhan\inst{1} \and
Zunian Wan\inst{1} \and
Yinqiang Zheng\inst{1}}

\authorrunning{J.~Zhao et al.}

\institute{The University of Tokyo}

\maketitle

\begin{abstract}

Image compression for machines calls for a unified codec that serves multiple downstream vision tasks. Existing approaches either adopt task-specific end-to-end designs, raising parameter and deployment overhead, or rely on transfer-based adaptations that remain externally attached and heuristic task design. 
A key limitation shared by both lines of work is their largely static computation pattern, which applies similar transformations across tokens despite the fact that different image regions exhibit markedly different semantic importance and complexity for machine perception.
We propose \method, a token-aware image compression framework that supports multiple downstream tasks within a single model. \method replaces the FFN layers in transformer-based compression model token-wise Mixture-of-Experts (MoE), enabling dynamic, token-level computation conditioned on the input content and task objective. To make MoE effective in compression model, we introduce a stable routing strategy that combines expert-choice routing with spatial total variation regularization to encourage spatially coherent assignments, and we propose a lightweight expert architecture, Group Shuffle MLP (GShMLP), to control parameter growth.
Extensive experiments show consistent improvement against baselines on both conventional image reconstruction and machine tasks.

\end{abstract}    
\section{Introduction}
\label{sec:intro}

\begin{figure}
    \centering
    \includegraphics[height=0.35\linewidth,width=0.85\linewidth]{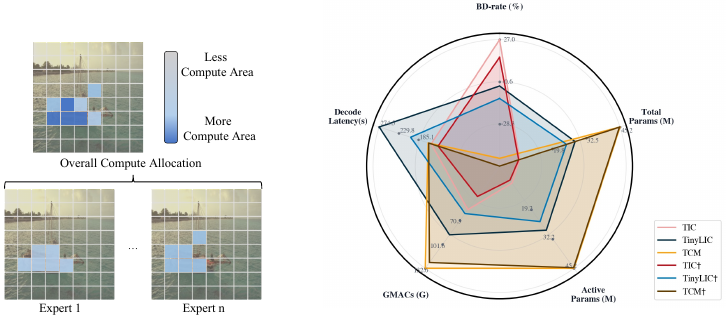}
    \caption{\textbf{Overview of dynamic compute allocation in \method.}
\textbf{Left:} Tokens are dispatched to different experts conditioned on the optimization objective, enabling heterogeneous compute allocation that focuses more computation on task-relevant regions.
\textbf{Right:} Reconstruction performance (BD-Rate) and efficiency (parameter efficiency and latency) on three transformer-based LIC backbones after applying \method.
Activated parameters denote the average number of parameters activated per token during a forward pass; detailed computation is provided in the Appendix.}
    \label{fig:teaser}
\end{figure}

Machine-oriented image compression has emerged as an active research area, driven by the rapid growth of machine-consumed visual data across a wide range of computer vision applications, such as classification, detection, and segmentation. 
However, directly applying image compression models optimized for human visual quality to machine vision tasks often removes task-relevant semantics and degrades downstream performance.
This has motivated a growing body of work aimed at developing unified compression frameworks that jointly optimize for multiple downstream machine vision. 
A straightforward strategy is task-specific end-to-end optimization, either with dedicated encoder-decoder pairs~\cite{le2021image,liu2022improving, song2021variable,wang2022deep} or with multi-branch task-adaptive architectures~\cite{agustsson2023multi,iwai2024controlling,zhang2024all,chamain2021end,duan2023unified,feng2022image}.
This design usually incurs substantial parameter overhead and deployment complexity.
Another line of work uses transfer-based adaptation, where a human-oriented base codec is adapted to each downstream task with a small set of trainable parameters~\cite{liu2023icmh,chen2023transtic,li2024image}.
Although parameter-efficient, these methods still depend on task-specific heuristics, and adaptation remains externally attached instead of being intrinsically learned in one unified model.

To overcome these limitations, we introduce \method, a unified image compression framework that incorporates token-wise Mixture-of-Experts (MoE) into transformer-based LIC methods to support joint optimization across multiple perceptual tasks.
Rather than relying on externally attached task-specific modules, \method replaces the standard FFN layers in transformer blocks with MoE layers, enabling dynamic token-level computation. Each input token is adaptively routed to specialized experts according to its content and task objective.
In contrast to prior methods that apply uniform transformations to all tokens regardless of perceptual demands, \method establishes a dynamic computation paradigm in which heterogeneous task objectives are accommodated intrinsically within a single unified model.

However, naively incorporating MoE into LIC models leads to sub-optimal performance. Through empirical analysis, we identify two key challenges.
First, many transformer-based LIC models employ point-wise tokenization (e.g., patch size of $1\times1$), where each token corresponds to an isolated spatial location. As a result, token-wise MoE routing decisions are made without spatial context, producing noisy and fragmented expert assignment patterns that manifest as salt-and-pepper artifacts in the routing maps. 
Second, FFN layers account for the majority of parameters in transformer-based LIC models. Directly replacing each FFN with $E$ independent expert networks therefore introduces an almost $E{\times}$ parameter overhead. Simply reducing the expert network size, however, degrades the channel aggregation capability of the FFN, leading to noticeable RD performance drops.
To address the spatially inconsistent routing, we adopt expert-choice (EC) routing~\cite{zhou2022mixture}, where each expert actively selects its preferred tokens from the full sequence rather than receiving independent per-token assignments. This design promotes balanced expert utilization and produces more spatially coherent routing decisions, as experts can leverage global context when selecting tokens. We further introduce a Spatial Total Variance Regularization that explicitly encourages piecewise-smooth expert assignments by penalizing high-frequency variations in the spatial expert affinity maps.
To alleviate parameter overhead, we propose Group Shuffle MLP (GShMLP), a lightweight expert architecture with two-layers grouped projections. To restore the cross-channel interaction lost due to grouping, we incorporate a parameter-free channel shuffle~\cite{zhang2018shufflenet} between the two projections. By setting the group number $G$, the total parameter count of \method can be reduced to the corresponding baseline with moderate performance degradation.

To summarize, our contributions are as follows:
\begin{itemize}
    \item We propose \method, a unified LIC framework that integrates token-wise Mixture-of-Experts into transformer-based codecs, enabling a single model to support both reconstruction quality and machine downstream tasks.

    \item We address two key challenges when introducing MoE into LIC: spatially discontinuous routing and parameter overhead. Specifically, we improve routing coherence through expert-choice routing with Spatial Total Variance Regularization, and reduce expert complexity through Group Shuffle MLP.

    \item Extensive experiments validate the effectiveness of \method. On reconstruction, \method improves BD-Rate by 11.54\% (TIC), 8.12\% (TinyLIC), and 6.42\% (TCM). On downstream tasks with TIC as the base codec, \method further improves BD-Rate by 7.74\% (classification), 8.98\% (detection), and 9.72\% (instance segmentation) against Full Finetune-TIC.
\end{itemize}

\section{Related Works}
\label{sec:rela}

\subsection{Learned Image Compression}
Learned Image Compression (LIC) was first introduced in~\cite{balle2017endtoendoptimizedimagecompression}, which adopts an autoencoder-based architecture to perform transform coding in the pixel space. Due to its superior rate-distortion (R-D) performance, LIC has shown great potential as a promising alternative to traditional image compression paradigms. A typical LIC codec consists of three key components: an analysis transform that maps the input image from the high-dimensional pixel space to a compact latent representation; an entropy model that encodes the latent variables into a compressed bitstream; and a synthesis transform that reconstructs the image from the latent space back to the pixel domain. 
The main research directions in LIC can be broadly categorized into two types. The first focuses on designing more efficient and expressive codec architectures, including more representative analysis and synthesis transforms. These structures have evolved from early CNN-based~\cite{balle2017endtoendoptimizedimagecompression,balle2018variationalimagecompressionscale,cui2022asymmetricgaineddeepimage,cheng2020learnedimagecompressiondiscretized} designs to Transformer-based~\cite{lu2021transformerbasedimagecompression,li2024frequencyawaretransformerlearnedimage,zou2022devildetailswindowbasedattention,liu2023learnedimagecompressionmixed} models, enabling better modeling capacity. In addition, recent efforts explore user-controllable compression, such as variable-rate coding~\cite{yang2022slimmablecompressiveautoencoderspractical,choi2019variable,li2024onceforallcontrollablegenerativeimage} and distortion--perception~\cite{agustsson2023multirealismimagecompressionconditional} trade-off control, to enhance flexibility in practical applications. 
The second line of research focuses on designing more powerful entropy models to better estimate the probability distribution of latent representations. This has evolved from factorized~\cite{balle2017endtoendoptimizedimagecompression} and hyperprior-based~\cite{balle2018variationalimagecompressionscale} models to more advanced autoregressive entropy models~\cite{lee2019contextadaptiveentropymodelendtoend,he2021checkerboardcontextmodelefficient,qian2022entroformertransformerbasedentropymodel,minnen2020channelwiseautoregressiveentropymodels,minnen2018jointautoregressivehierarchicalpriors}.
However, most LIC methods are human-centric, as they are typically optimized using perceptual quality metrics such as MSE or MS-SSIM. While effective for human viewing, such objectives may not align with the needs of machine vision. In particular, pixel-wise distortion metrics tend to over-allocate bits to visually fine-grained details, while potentially neglecting semantically important structures that are critical for downstream vision tasks.

\subsection{Multi-Task Image Compression}

MT-IC has been extensively studied~\cite{yan2021sssic,song2021variable,li2024human,choi2022scalable,chamain2021end,duan2023unified,feng2022image}. Early work focused on the trade-off between reconstruction fidelity and human perception~\cite{agustsson2023multi,iwai2024controlling,zhang2024all}, but the growing prevalence of vision models has shifted attention toward coding for machines~\cite{li2024human,chen2023transtic,duan2023unified,feng2022image}. 
A straightforward strategy is per-task customization---training a dedicated encoder--decoder for each task~\cite{chamain2021end,le2021image,liu2022improving,song2021variable,wang2022deep}---yet this inflates parameters and training cost. To reduce duplication, multi-branch codecs introduce task-specific pathways at the decoder side, including separate/multi-path~\cite{zhang2024all,song2021variable}, or conditional decoders~\cite{agustsson2023multi,iwai2024controlling,zhang2024all} and scalable bottleneck $\hat{\yb}$~\cite{choi2022scalable,harell2022rate,hu2020towards,yan2021sssic,wang2021towards} tailored to each task. 
These designs typically adopt a unified encoder and entropy model and derive a shared bottleneck feature \(\hat{\yb}\) for all tasks (i.e., a single bitstream), but still activate a large number of parameters.
A complementary line adopts task-specific PEFT~\cite{chen2023transtic,li2024image,liu2023icmh}, inserting lightweight adapters or prompt layers into a shared codec to enable per-task adaptation with minimal parameters. 
Despite these advances, prior MT-IC systems largely perform image-wide, uniform computation and lack token-level task discrimination. \method addresses this gap by replacing dense FFNs with sparsely activated MoE layers and coupling routing with task prompts, thereby introducing token-level granularity for heterogeneous computation in MT-IC and enabling precise task specialization.

\subsection{Mixture of Experts (MoE)}
Mixture-of-Experts (MoE)~\cite{shazeer2017outrageously} is a representative form of conditional computation~\cite{abati2020conditional, cho2014exponentially,lin2019conditional,puigcerver2020scalable}, where computation can be dynamically allocated based on the input. It has been extensively utilized to scale model capacity while keeping the increase in computation close to that of the original dense baseline. MoE has been comprehensively demonstrated to be successful in large-scale language models~\cite{liu2024deepseek,li2025minimax,zoph2202st}, and vision-models~\cite{yu2022scaling,fei2024scaling,xue2023raphael,park2023denoising,park2024switch,lee2024multi,feng2023ernie,balaji2022ediff}.
Most existing MoE architectures adopt a token-choice (TC) routing strategy, where each token independently selects top-$k$ experts for processing~\cite{du2022glam,fedus2022switch,lepikhin2020gshard}. However, TC often suffers from load imbalance: a small subset of experts is over-utilized while many others remain underused, so auxiliary load-balancing losses are typically required to regularize the routing~\cite{zhenxing2022switch, sun2024ec,shi2025diffmoe}. To address this,~\cite{zhou2022mixture} propose an alternative Expert-Choice (EC) routing scheme, in which, instead of tokens choosing experts, each expert selects its top-$C$ tokens to process. EC inherently ensures balanced utilization of all experts, and enables each expert get access to all tokens within an entire image. 
Inspired by these applications, we introduce MoE architecture into the domain of image coding, building a codec with token-level computational granularity, and demonstrating the effectiveness of this sparsely gated, adaptive compute allocation strategy in multi-task coding.

\section{Method}
\label{sec:method}

\begin{figure*}
    \centering
    \includegraphics[height=0.48\textwidth,width=0.95\textwidth]{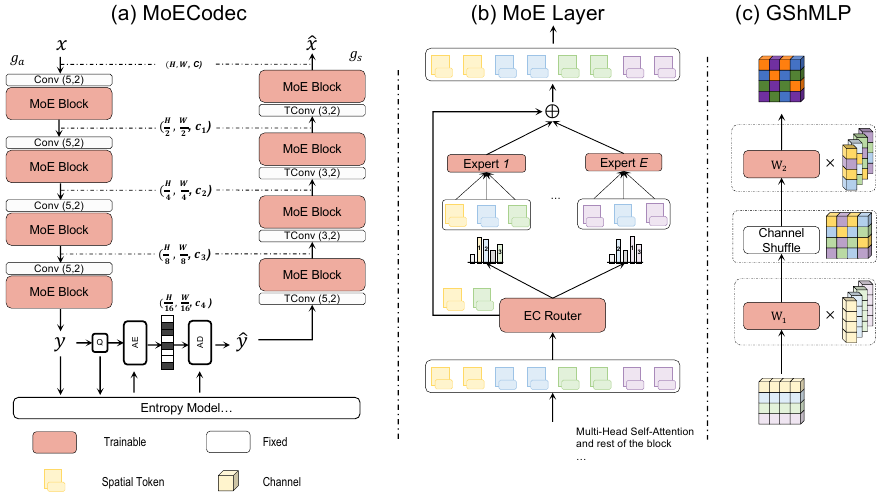}
    \caption{\texttt{(a)}: Overall architecture of \method. Starting from a pre-trained Transformer-based base codec, all transformer blocks in the encoder and decoder are replaced with MoE blocks while keeping other components fixed. \texttt{(b)}: Structure of a MoE layer in each MoE block. Input tokens are distributed to different experts through EC routing, enabling token-adaptive computation within the block. \texttt{(c)}: Each expert network is instantiated as the proposed Group Shuffle MLP.}
    \label{fig:overview}
\end{figure*}

\subsection{Unified Compression via MoE}
\noindent\textbf{{Preliminaries.}}
A typical LIC model~\cite{balle2017endtoendoptimizedimagecompression} adopts an autoencoder architecture, composed of an analysis 
encoder $g_a$ and a synthesis decoder $g_s$. The encoder $g_a$ maps the input image $\xb$ from the high-dimensional pixel space to a compact latent representation $\yb = g_a(\xb)$, which is subsequently quantized to $\hat{\yb}$. The decoder $g_s$ then reconstructs the image from the quantized latent, i.e., $\hat{\xb} = g_s(\hat{\yb})$.
Directly storing $\hat{\yb}$ would incur significant storage overhead. To address this, 
LIC models the distribution of $\hat{\yb}$ via a learned entropy model $p(\hat{\yb})$, 
enabling efficient entropy coding. The compression objective balances rate and distortion, 
formulated as:
\begin{equation}
    \mathcal{L}_\text{rd} = \mathbb{E}_{\xb \sim p_\xb}[\lambda_{rd}\cdot r(\hat{\yb}) +  \Db(\xb,\hat{\xb})]
    \label{equ:rd}
\end{equation}
where $r(\hat{\yb})$ denotes the estimated bitrate, $\Db(\xb, \hat{\xb})$ is a distortion metric and $\lambda_{rd}$ is a hyperparameter that controls the rate-distortion trade-off.

\noindent{\textbf{\method.}}
Standard LIC models apply computation uniformly across all image regions, regardless of local complexity. This static computation paradigm lacks fine-grained, dynamic adaptability and fails to provide sufficient capacity to support heterogeneous perceptual objectives within a single unified model. To this end, we propose \method, which replaces the standard FFN layers with Mixture-of-Experts (MoE) layers within the transformer blocks of both encoder and decoder (see~\Cref{fig:overview}), enabling dynamic, content-adaptive computation paths across the encoding and decoding process.
Each MoE layer in \method consists of a learned routing network and $E$ parallel expert networks. The routing network takes the intermediate feature $\Xb \in \mathbb{R}^{B \times S \times d}$ as input and determines the token-expert assignment, dispatching each token to one or more expert networks for processing. Each expert network is instantiated as a lightweight two-layer Group Shuffle MLP (GShMLP), which partitions the input channels into groups and applies a lightweight nonlinear transform within each group independently. An explicit channel shuffle is inserted between the two grouped linear layers to facilitate cross-group channel aggregation.

\subsection{Routing Strategy}

\textbf{Expert-Choice Routing.}
The routing strategy plays a critical role in MoE layers, as it specifies how tokens are assigned to experts and thus determines the model's computational allocation pattern.
A common baseline is token-choice (TC) routing, which dispatches each token independently to its top-$k$ preferred experts. While TC routing is generally effective, its fully independent token-wise decision mechanism becomes less aligned with LIC models, which typically adopt extremely fine-grained tokenization (e.g., patch size $1{\times}1$ at the pixel/point level). 
Under such dense spatial sampling, routing decisions are made at isolated spatial locations, often leading to noisy and spatially fragmented expert assignment maps.
To alleviate this issue, we first adopt expert-choice (EC) routing~\cite{zhou2022mixture}, in which each expert actively selects its top-$k$ tokens from the entire token sequence. By reversing the routing perspective from token-wise local decisions to expert-wise global selection, EC routing enables experts to consider global token competition during assignment. 
This global view mitigates locally inconsistent routing decisions and naturally encourages more coherent expert utilization across spatial regions. 
Moreover, EC routing enforces a fixed token budget per expert by construction, ensuring balanced expert utilization without requiring additional auxiliary load-balancing losses.

Formally, given an input token sequence 
$\Xb \in \mathbb{R}^{B \times S \times d}$, 
the router computes a token--expert affinity matrix via a learned projection 
$\Wb_r \in \mathbb{R}^{d \times E}$:
\begin{equation}
    \Ab = \mathrm{softmax}(\Xb \Wb_r, \mathrm{dim}{=}-1)
    \in \mathbb{R}^{B \times S \times E}.
    \label{equ:A}
\end{equation}

For each batch $b$ and expert $e$, 
we select the top-$k$ tokens along the sequence dimension:
\begin{equation}
    \Ib_{b,e} = 
    \mathrm{TopK}\!\left(\Ab_{b,:,e},\, k\right),
\end{equation}
where $k = \lfloor S \cdot f_c / E \rfloor$ and $f_c$ is the capacity factor.

The gating tensor $\Gb \in \mathbb{R}^{B \times S \times E}$ 
retains affinity scores for selected tokens and zeros out the rest:
\begin{equation}
    \Gb_{b,s,e} =
    \begin{cases}
        \Ab_{b,s,e}, & \text{if } s \in \Ib_{b,e}, \\
        0, & \text{otherwise.}
    \end{cases}
\end{equation}

The output is computed by aggregating expert outputs:
\begin{equation}
    \Xb^{\mathrm{out}}_{b,s} 
= \frac{\sum_{e=1}^{E} \Gb_{b,s,e}\,\mathcal{F}_e(\Xb_{b,s})}
{\sum_{e=1}^{E}\Gb_{b,s,e} + \epsilon},
\end{equation}
where $\epsilon$ is a small constant for numerical stability.

\textbf{Spatial Total Variance Regularization.}
Although EC routing promotes globally balanced expert utilization, routing decisions are still made at the pixel level and may exhibit local spatial inconsistencies. 
To further encourage spatial coherence in expert assignment, we introduce a spatial total variance (TV) regularization on the token--expert affinity maps.

Specifically, we reshape the affinity tensor 
$\Ab \in \mathbb{R}^{B \times S \times E}$ 
into a 2D spatial representation 
$\tilde{\Ab} \in \mathbb{R}^{B \times E \times H \times W}$, 
where $S = H \times W$. 
We then apply an anisotropic total variation (TV) regularization over spatial dimensions:
\begin{equation}
    \mathcal{L}_{\mathrm{r}} = 
    \frac{1}{B E}\sum_{b=1}^{B}\sum_{e=1}^{E}
    \left(
        \|\nabla_h \tilde{\Ab}_{b,e}\|_1 +
        \|\nabla_w \tilde{\Ab}_{b,e}\|_1
    \right),
\end{equation}
where $\nabla_h$ and $\nabla_w$ denote differences along 
the height and width dimensions, respectively.
This regularization encourages spatial smoothness in expert affinity maps, promoting coherent expert specialization across neighboring tokens.
It is applied only during training and introduces no additional inference overhead.

In summary, the overall training objective is:
\begin{equation}
    \mathcal{L}_{\mathrm{total}} 
    = \mathcal{L}_{\mathrm{rd}} 
    + \alpha \mathcal{L}_{\mathrm{r}},
    \label{equ:alpha}
\end{equation}
where $\alpha$ controls the strength of spatial regularization.

\subsection{Group Shuffle MLP}
\label{sec:GroupShMLP}

Each expert $\mathcal{F}_e$ in the MoE layer is instantiated as a Group Shuffle MLP (GShMLP), a lightweight expert architecture designed to reduce parameter complexity while preserving effective cross-channel interaction.

Standard FFN layers consist of two linear projections with an expansion ratio of 4:
\begin{equation}
    \mathrm{FFN}(\Xb) = \sigma(\Xb\Wb_1)\Wb_2,
    \quad \Wb_1 \in \mathbb{R}^{d \times 4d},\;
    \Wb_2 \in \mathbb{R}^{4d \times d},
\end{equation}
In a naive MoE design, each of the $E$ experts is an independent FFN, yielding $E \cdot 8d^2$ parameters in total---scaling linearly with $E$. A straightforward remedy, i.e., reducing the expansion ratio, degrades the channel aggregation capacity of the FFN, leading to a noticeable RD performance drop (see~\Cref{tab:ablation}). Instead, GShMLP reduces parameter complexity through grouped channel processing, while restoring cross-group information exchange via a parameter-free channel shuffle~\cite{zhang2018shufflenet}.

Formally, GShMLP retains the two-projection structure of a standard FFN,
but constrains the projection matrices to block-diagonal form.
Specifically, we decompose the channel dimension into $G$ groups
and define:

\begin{equation}
    \Wb_1 = \mathrm{diag}(\Wb_1^{(1)}, \dots, \Wb_1^{(G)}) 
    \in \mathbb{R}^{d \times 4d}, \quad \Wb_1^{(g)} \in \mathbb{R}^{\frac{d}{G} \times \frac{4d}{G}}
\end{equation}

\begin{equation}
    \Wb_2 = \mathrm{diag}(\Wb_2^{(1)}, \dots, \Wb_2^{(G)}) 
    \in \mathbb{R}^{4d \times d}, \quad \Wb_2^{(g)} \in \mathbb{R}^{\frac{4d}{G} \times \frac{d}{G}}
\end{equation}
A parameter-free channel shuffle is inserted between the two projections to enable cross-group interaction:
\begin{equation}
    \mathrm{GShMLP}(\Xb) 
    = \left(\mathrm{Shuffle}\!\left(\sigma(\Xb \Wb_1)\right)\right)\Wb_2.
\end{equation}


\section{Experiments}
\label{sec:experiments}
To comprehensively evaluate \method, we conduct experiments from three perspectives.
First, to assess its performance on conventional image reconstruction, 
we compare \method with three transformer-based LIC baselines: 
TIC~\cite{lu2021transformerbasedimagecompression}, 
TinyLIC~\cite{ma2024tinylichighefficiencylossyimage}, 
and TCM~\cite{liu2023learned}.
Second, to evaluate its adaptability to machine-oriented tasks, 
we adopt TIC as the base codec and compare \method against strong transfer-based baselines, including full fine-tuning, AdaptICMH-TIC~\cite{liu2023icmh}, and TransTIC~\cite{chen2023transtic}.
Finally, to better understand the design choices of \method, 
we conduct extensive ablation studies analyzing the effects of routing strategy, GShMLP, the number of experts, and the placement of MoE layers.

\subsection{Experimental Setup}
\label{sec:setup}

\noindent\textbf{Training and Datasets.}
Unless otherwise specified, we use $E{=}4$ experts and $G{=}8$ groups, with capacity factor $f_c{=}1.0$ (Eq.~\ref{equ:A}) and spatial regularization weight $\alpha{=}1\times10^{-3}$ (Eq.~\ref{equ:alpha}).
This default setting is selected to match the parameter budget of baseline TIC.
For conventional reconstruction, we train from scratch on Flickr2W under both MSE and MS-SSIM objectives for 3M iterations.
We use Adam with initial learning rate $1\times10^{-4}$, decayed to $1\times10^{-5}$ in the final 25\% iterations.
For machine-oriented tasks, we use TIC initialized by pre-trained weights from~\cite{chen2023transtic} and train only router/expert parameters, while freezing the remaining codec parameters.
For image classification, we train on ImageNet-1K~\cite{deng2009imagenet} for 8 epochs with batch size 16.
For object detection and instance segmentation, we train on COCO 2017 train~\cite{lin2014microsoft} for 40 epochs with batch size 8.
To construct RD curves, we use $\lambda_{\mathrm{rd}} \in \{0.0005, 0.001, 0.002, 0.005, 0.007, 0.01\}$.
For transfer baselines (TransTIC and AdaptICMH), we follow their original training protocols and report results under the same evaluation pipeline as ours.

\noindent\textbf{Evaluation.}
For reconstruction quality, we report PSNR and MS-SSIM.
Evaluations are conducted on the Kodak and CLIC~\cite{toderici2020workshop} datasets. 
For classification, we evaluate on the ImageNet validation set with resize + center crop to $256 \times 256$.
We use pretrained ResNet50 from torchvision as the off-the-shelf evaluator and report top-1 accuracy.
For object detection and instance segmentation, we evaluate on the COCO 2017 validation set.
We use pre-trained Faster R-CNN and Mask R-CNN from Detectron2 as evaluators, respectively.
All test images are resized such that the shorter side is 800 pixels.
We report mAP at IoU=0.5.
\subsection{Results of multi-task performance}
\label{sec:mul_task}
\textbf{Multi-task Performance.}
We evaluate \method from both reconstruction and machine-task perspectives.
For reconstruction, we apply \method to three transformer LIC backbones (TIC~\cite{lu2021transformerbasedimagecompression}, TinyLIC~\cite{ma2024tinylichighefficiencylossyimage}, and TCM~\cite{liu2023learned}) and report BD-Rate on Kodak with VTM-17.1 as anchor.
For machine-oriented evaluation, we use TIC as the base codec and compare against Full Finetune-TIC, TransTIC~\cite{chen2023transtic}, and AdaptICMH-TIC~\cite{liu2023icmh} on classification, object detection, and instance segmentation.
Unless otherwise stated, we use \method with $E{=}4$ and $G{=}8$; this $E/G$ setting is chosen to match the parameter budget of baseline TIC.

\begin{figure*}
    \centering
    \includegraphics[height=0.55\textwidth,width=1.0\textwidth]{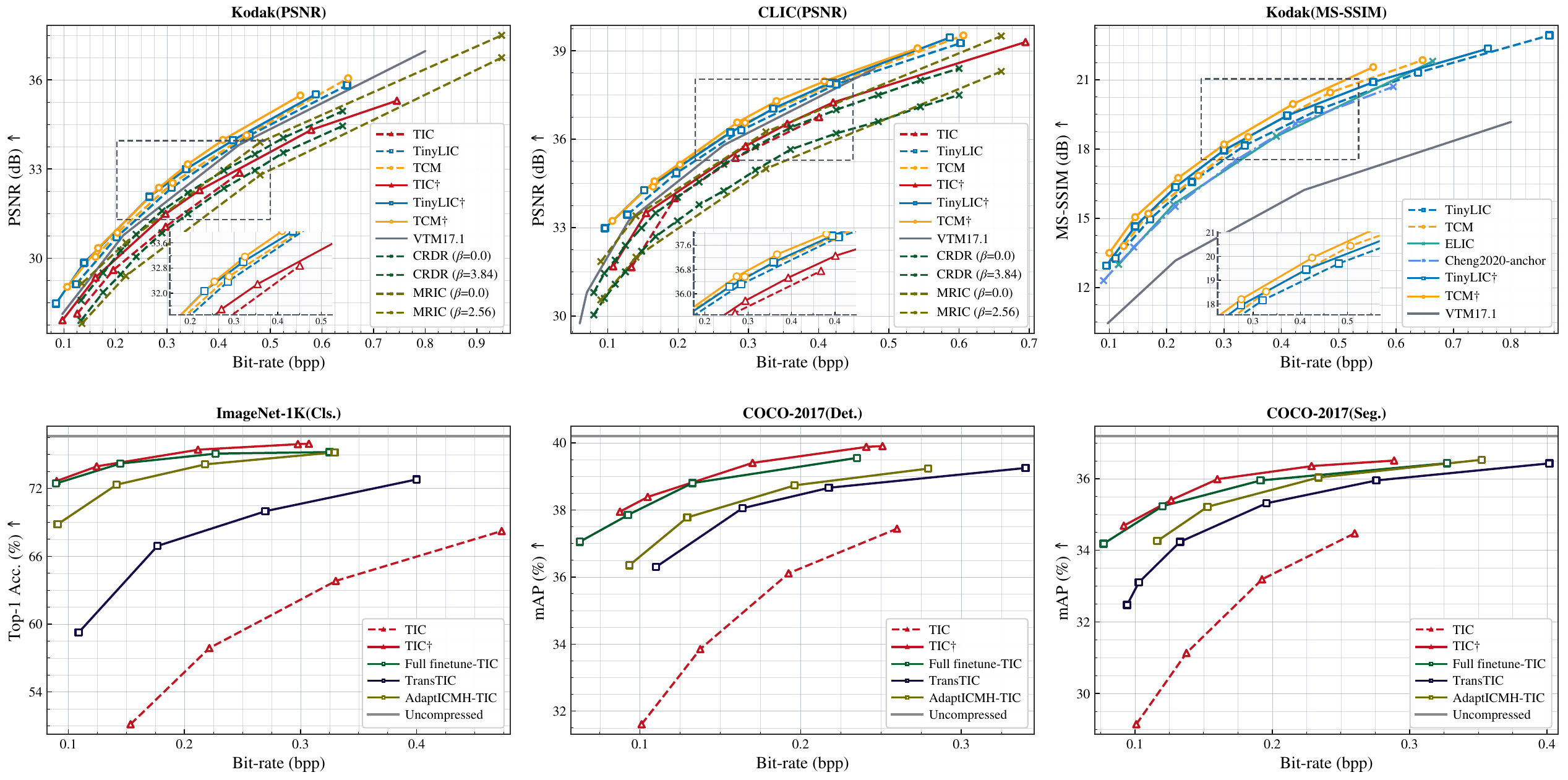}
    \caption{Multi-task performance. Models applied using \method is marked by $\dagger$.}
    \label{fig:rd}
\end{figure*}

\Cref{tab:main_results} shows that, on reconstruction (PSNR, Kodak), applying \method to TIC, TinyLIC, and TCM yields BD-Rate improvements of 11.54\%, 8.12\%, and 6.42\%, respectively, while reducing total parameters by 4.5\%, 11.4\%, and 0.44\%.
For downstream tasks (anchor: Full finetune-TIC), \method achieves BD-Rate gains of 7.74\% on classification, 8.98\% on detection, and 9.72\% on instance segmentation.
At the same time, activated parameters are reduced from 7.51M to 6.25M, showing improved utility-efficiency trade-off (See Appendix for the calculation method of activated parameters.)
These trends are consistent with the curve-level comparison in \Cref{fig:rd}.
The results demonstrate that \method provides a consistent improvement over strong baselines in both reconstruction quality and machine-task utility under constrained model cost.
Next, we perform controlled ablations to isolate the contribution of each design choice.

\begin{table*}[t]
\renewcommand{\arraystretch}{0.8}
\setlength{\abovecaptionskip}{0pt}
\centering
\caption{Multi-task performance comparison. BD-Rate (\%) is computed 
against VTM-17.1 and Full Finetune, respectively. $\dagger$ denotes \method applied to the corresponding backbone. \textcolor{red}{Red} and \textcolor{blue}{Blue} values denote positive and negative changes relative to the corresponding backbone baseline. GMACs are calculated on $768\times512$ input images.}
\resizebox{.90\linewidth}{!}{
\begin{tabular}{@{}l|cc|c|c|c|c@{}}
\toprule
\multirow{2}{*}{\textbf{Method}} 
  & \multicolumn{2}{c|}{\textbf{GMACs}} 
  & \textbf{\#Params}
  & \textbf{\#Train.Params}
  & \textbf{\#Act.Params}
  & \textbf{BD-Rate} \\
\cmidrule(lr){2-3}
  & \textbf{Enc.} & \textbf{Dec.}
  & (M)
  & (M)
  & (M)
  & (\%) \\
\midrule
\multicolumn{7}{l}{\footnotesize\textit{PSNR (Kodak)}} \\
\midrule
TIC~\cite{lu2021transformerbasedimagecompression}              &  28.4 & 28.2 & 7.51 & 7.51 & 7.51 & 26.98                 \\
TIC$^\dagger$            & 19.8 & 19.7 & 7.17\textcolor{red}{(-4.5\%)}  & 7.17 & 6.25 &15.44\textcolor{red}{(-11.54)} \\
\midrule
TinyLIC~\cite{ma2024tinylichighefficiencylossyimage}          & 45.0 & 44.8 & 28.34 & 28.34 & 28.34 & -3.42                 \\
TinyLIC$^\dagger$        & 31.0 & 30.7 & 25.11\textcolor{red}{(-11.4\%)}  & 25.11 & 24.54 &-11.54 \textcolor{red}{(-8.12)} \\
\midrule
TCM~\cite{liu2023learnedimagecompressionmixed}        & 57.4 & 75.3 & 45.2 & 45.2 & 45.18  & -5.45                 \\
TCM$^\dagger$            & 53.6 & 71.5 & 45.0\textcolor{red}{(-0.44\%)} & 45.0 & 44.93 &-11.87\textcolor{red}{(-6.42)} \\
\midrule
\multicolumn{7}{l}{\footnotesize\textit{MS-SSIM (Kodak)}} \\
\midrule
TinyLIC~\cite{ma2024tinylichighefficiencylossyimage}        & 45.0 & 44.8 & 28.34 & 28.34 & 28.34 & -48.60                 \\
TinyLIC$^\dagger$        & 31.0 & 30.7 & 25.11\textcolor{red}{(-11.4\%)}  & 25.11 & 24.54 &-52.58\textcolor{red}{(-3.98)} \\
\midrule
TCM~\cite{liu2023learnedimagecompressionmixed}       & 57.4 & 75.3 & 45.2 & 45.2 & 44.93 & -50.82                 \\
TCM$^\dagger$            & 53.6 & 71.5 & 45.0\textcolor{red}{(-0.44\%)} & 45.0 & 44.93 &-55.90\textcolor{red}{(-5.08)} \\
\midrule
\multicolumn{7}{l}{\footnotesize\textit{Classification (ImageNet-1K-val)}} \\
\midrule
Full finetune TIC  & 28.4 & 28.2 & 7.51 & 7.51 & 7.51 & --                 \\
TransTIC~\cite{chen2023transtic}             & 65.6 & 39.9 &9.12\textcolor{blue}{(+21.4\%)} & 1.61 & 9.12 & 35.77                 \\
AdaptICMH~\cite{liu2023icmh}            & 31.2 & 31.1 & 7.80\textcolor{blue}{(+3.8\%)}  & 0.29 & 7.80 & 5.01                 \\
AdaptICMH-Large~\cite{liu2023icmh}            & 65.0 & 75.9 & 8.95\textcolor{blue}{(+19.2\%)} & 1.44 & 8.95 & 4.7                 \\
TIC$^\dagger$                & 19.8 & 19.7 & 7.17\textcolor{red}{(-4.5\%)} & 1.24 & 6.25 &\textbf{-7.74} \\
\midrule
\multicolumn{7}{l}{\footnotesize\textit{Object Detection (COCO2017-val)}} \\
\midrule
Full finetune TIC  & 28.4 & 28.2 & 7.51 & 7.51 & 7.51 & --                 \\
TransTIC~\cite{chen2023transtic}             & 65.6 & 39.9 & 9.12\textcolor{blue}{(+21.4\%)} & 1.61 & 9.12 & 26.09                 \\
AdaptICMH~\cite{liu2023icmh}            & 31.2 & 31.1 & 7.80\textcolor{blue}{(+3.8\%)} & 0.29 & 7.80 & 15.98                 \\
AdaptICMH-Large~\cite{liu2023icmh}            & 65.0 & 75.9 & 8.95\textcolor{blue}{(+19.2\%)} & 1.44 & 8.95 & 14.38                 \\
TIC$^\dagger$         & 19.8 & 19.7 & 7.17\textcolor{red}{(-4.5\%)} & 1.24 & 6.25 &\textbf{-8.98} \\
\midrule
\multicolumn{7}{l}{\footnotesize\textit{Instance Segmentation (COCO2017-val)}} \\
\midrule
Full finetune TIC  & 28.4 & 28.2 & 7.51 & 7.51 & 7.51 & --                 \\
TransTIC~\cite{chen2023transtic}             & 65.6 & 39.9 & 9.12\textcolor{blue}{(+21.4\%)} & 1.61 & 9.12 & 12.83                 \\
AdaptICMH~\cite{liu2023icmh}            & 31.2 & 31.1 & 7.80\textcolor{blue}{(+3.8\%)}  & 0.29 & 7.80 & 6.34                 \\
AdaptICMH-Large~\cite{liu2023icmh}            & 65.0 & 75.9 & 8.95\textcolor{blue}{(+19.2\%)} & 1.44 & 8.95 & 5.71                 \\
TIC$^\dagger$        & 19.8 & 19.7 & 7.17\textcolor{red}{(-4.5\%)} & 1.24 & 6.25 &\textbf{-9.72} \\
\bottomrule
\end{tabular}}
\label{tab:main_results}
\end{table*}

\subsection{Ablation Study}
\label{sec:ablation}
We perform controlled ablations on TIC and report Kodak PSNR BD-Rate (\%) against VTM-17.1.
Unless otherwise stated, all settings follow the default configuration and are trained with the same protocol.
Results are summarized in~\Cref{tab:ablation}. A more detailed results are in Appendix.

\noindent\textbf{Routing strategy.}
We compare three routing variants: token-choice (TC) routing, expert-choice (EC) routing, and EC with spatial regularization.
For TC routing, we follow~\cite{fedus2022switchtransformersscalingtrillion} by using top-1 routing with a capacity factor of 1 and an auxiliary load-balancing loss; we set its weight to $5\times 10^{-4}$ for all training bitrates.
For EC routing, we remove the spatial regularization term in~\Cref{equ:alpha}.
The resulting routing maps are visualized in~\Cref{fig:expert_1}. TC yields the noisiest and most fragmented assignments, while EC produces noticeably more spatially coherent maps; adding spatial regularization further improves the continuity of expert assignments.
These qualitative trends are consistent with quantitative results in~\Cref{tab:ablation}: replacing TC with EC improves BD-Rate from 19.57 to 15.90, and adding spatial regularization further reduces it to 15.44.

\noindent\textbf{Expert architecture.}
Using dense FFN experts (ratio=4) achieves the best BD-Rate (14.82), but increases parameters to 12.27M.
Reducing FFN expansion (ratio=1) lowers parameters to 7.53M but degrades BD-Rate to 19.07.
GShMLP obtains a better efficiency-performance trade-off: 15.44 BD-Rate at 7.17M parameters, substantially better than reduced FFN under a similar budget.

\noindent\textbf{Expert number.}
Intuitively, using more experts increases the diversity of available coding modes, potentially expanding the search space of content-adaptive transforms.
We therefore investigate the effect of the expert number $E$ on \method.
To keep the parameter count comparable to the baseline while scaling $E$, we set the grouping factor in GShMLP to $G=2E$.
As shown in~\Cref{tab:ablation}, increasing $E$ consistently improves rate--distortion performance: BD-Rate decreases from 16.02 to 15.44 and further to 14.59.

\noindent\textbf{MoE placement.}
We ablate where to insert MoE layers in \method by placing them in the encoder only, the decoder only, or both.
Under task-wise training, introducing MoE in either stage already enables task-dependent heterogeneous computation and yields around a 10\% BD-Rate gain over the baseline.
But placing MoE in both the encoder and decoder provides the most flexibility and achieves the best overall performance.

\begin{table}[t]
\centering
\caption{Ablation study on \method (Kodak, PSNR BD-Rate \% vs.\ VTM17.1).
The default \method configuration is marked with $\star$.}
\label{tab:ablation}
\setlength{\tabcolsep}{5pt}
\resizebox{.90\linewidth}{!}{
\begin{tabular}{llcccc}
\toprule
Group & Variant & Routing & Expert & Params (M) & BD-Rate $\downarrow$ \\
\midrule
\multirow{3}{*}{Routing}
  & TC routing          & \texttt{TC}  & \texttt{GShMLP} & 7.17 & 19.57  \\
  & EC routing          & \texttt{EC}  & \texttt{GShMLP} & 7.17 & 15.90  \\
  & EC + spatial reg$^\star$ & \texttt{EC} & \texttt{GShMLP} & 7.17 & \textbf{15.44} \\
\midrule
\multirow{3}{*}{Expert}
  & FFN (ratio=4)       & \texttt{EC}  & \texttt{FFN}    & 12.27 & \textbf{14.82}  \\
  & FFN (ratio=1)       & \texttt{EC}  & \texttt{FFN$'$} & 7.53 & 19.07 \\
  & GShMLP$^\star$      & \texttt{EC}  & \texttt{GShMLP} & 7.17 & 15.44 \\
\midrule
\multirow{3}{*}{$E,G$}
  & $E=2, G=4$               & \texttt{EC}  & \texttt{GShMLP} & 7.14 & 16.02  \\
  & $E=4, G=8$$^\star$       & \texttt{EC}  & \texttt{GShMLP} & 7.17 & 15.44  \\
  & $E=8, G=16$               & \texttt{EC}  & \texttt{GShMLP} & 7.23 & \textbf{14.59} \\
\midrule
\multirow{3}{*}{Placement}
  & Encoder only        & \texttt{EC}  & \texttt{GShMLP} & 7.34 &17.62  \\
  & Decoder only        & \texttt{EC}  & \texttt{GShMLP} & 7.34 & 17.54 \\
  & Enc.+Dec.$^\star$   & \texttt{EC}  & \texttt{GShMLP} & 7.17 & \textbf{15.44} \\
\bottomrule
\end{tabular}
}
\end{table}
\vspace{-3mm}

\subsection{Qualitative Results}
\label{sec:qualitative}
\begin{figure*}[t]
    \centering
    \includegraphics[height=0.95\linewidth,width=1.0\linewidth]{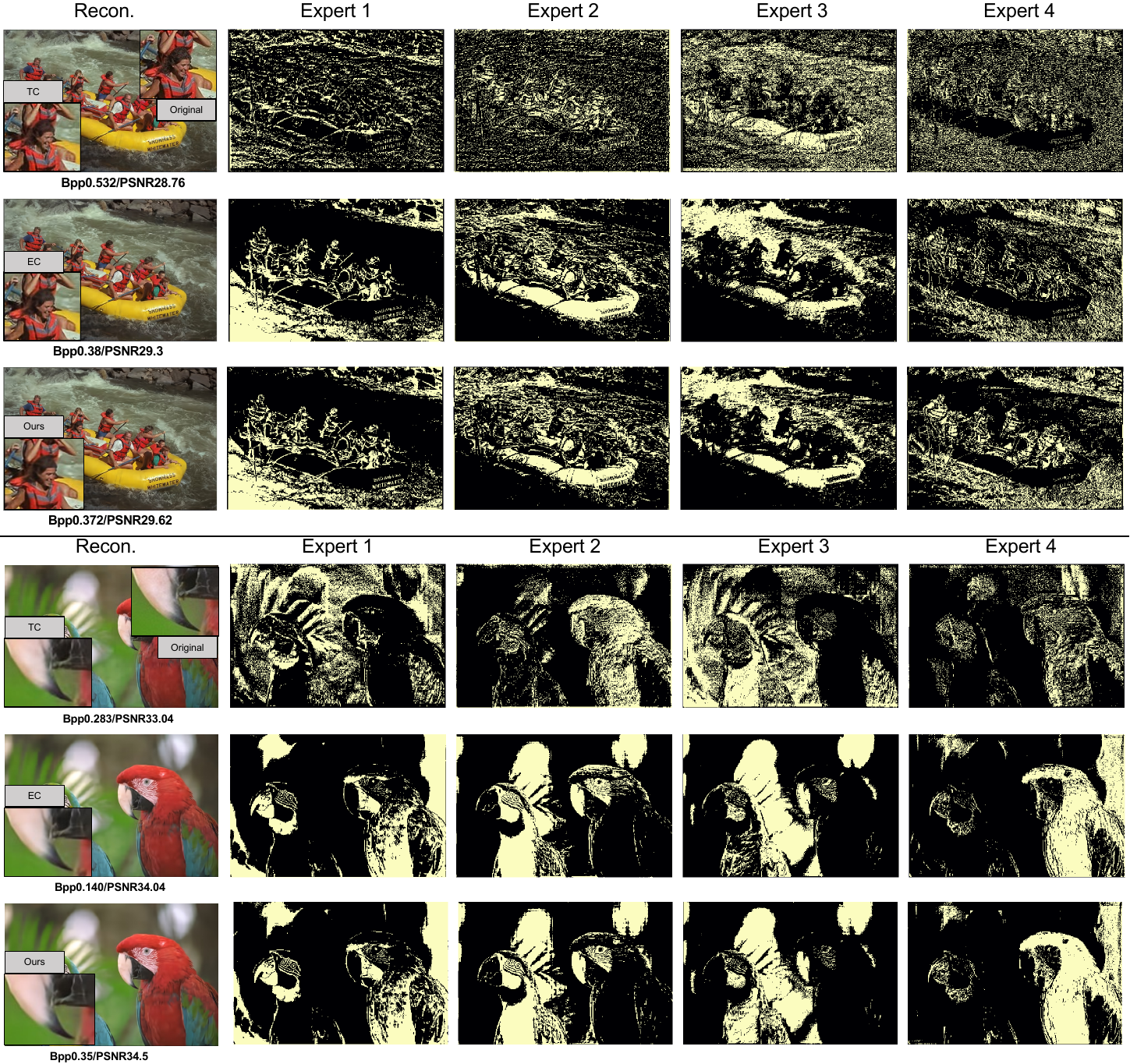}
    \caption{\textbf{Expert specialization} comparison under TC, EC, and EC+Spatial Total Variance regularization. Images drown from \texttt{Kodak14}, \texttt{Kodak23}. Routing results at encoder stage $g_{a1}$. EC-based routing yields more balanced and spatially coherent expert allocation than TC, and Spatial regularization further refines continuity.}
    \label{fig:expert_1}
\end{figure*}

\begin{figure}[!t]
    \centering
    \includegraphics[height=1.0\linewidth,width=0.95\linewidth]{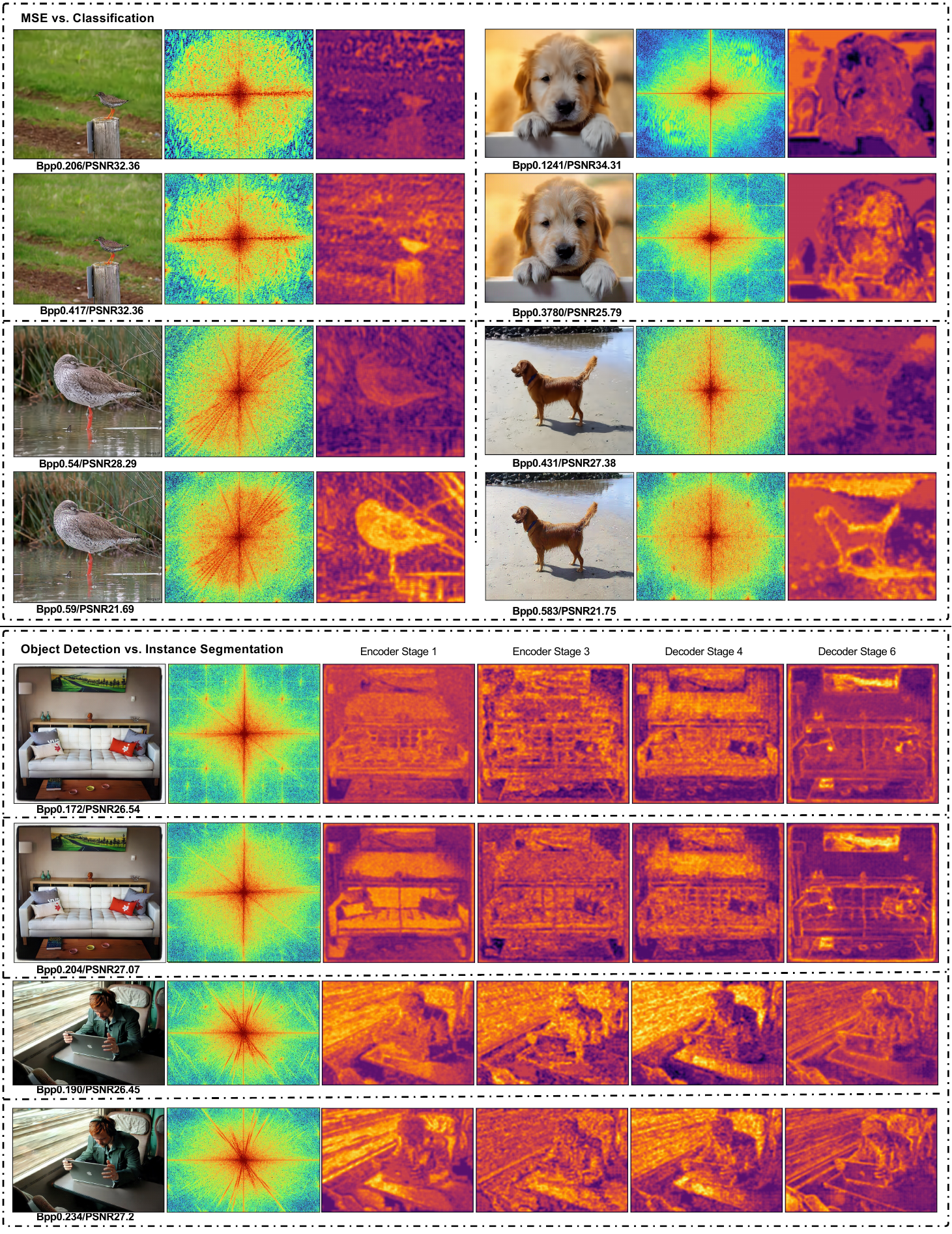}
    \caption{\textbf{Expert-token allocation across tasks.}
    Top: four samples from ImageNet validation coded for MSE (upper row) vs.\ classification (lower row); columns show reconstructed image, FFT spectrum, and the stage-$g_{a1}$ heatmap.
    Bottom: two samples from COCO2017 validation coded for object detection (upper row) vs.\ instance segmentation (lower row); columns show reconstructed image, FFT spectrum, and stage-wise heatmaps.
    Brighter regions indicate higher compute allocation (more experts selected).}
    \label{fig:expert_map}
\end{figure}

\noindent\textbf{Expert Specialization.}
We visualize expert specialization by plotting per-expert token-selection maps from the inter-slice features at the first encoder stage.
For each expert, selected tokens are highlighted, and three routing strategies are compared: token-choice (TC), expert-choice (EC), and EC with Spatial Total Variance regularization.
As shown in~\Cref{fig:expert_1}, TC produces noisy and spatially fragmented expert-token maps, and also shows expert imbalance (e.g., on \texttt{Kodak14}, Expert 3 is over-activated and receives more tokens than the other experts).
Switching to EC markedly improves spatial coherence and yields clearer functional specialization.
For instance, on \texttt{Kodak14}, Experts 1 and 4 mainly focus on the green parrot, while Experts 2 and 3 focus more on the red parrot.
A mild discontinuity is still observable under EC, and is further reduced by adding Spatial Total Variance regularization (third row).
Consistently, reconstruction quality improves from TC to EC and further to EC+Spatial TV.

\noindent\textbf{Compute Allocation Visualization.}
To analyze task-dependent compute behavior, we compare token-expert selection patterns for the same image under different task objectives.
Specifically, we visualize heatmaps of how many experts select each token.
As shown in~\Cref{fig:expert_map}, \method exhibits clear task-adaptive allocation.
\texttt{MSE vs. Classification:}
When optimized for classification, \method allocates more compute to semantically relevant targets and mid/low-frequency structures.
For large objects (e.g., the 2nd and 3rd examples), higher allocation concentrates on object texture regions; for small objects (e.g., the 1st and 4th examples), allocation focuses more on object contours and edges.
In contrast, under MSE optimization, \method tends to allocate more compute to strong high-frequency variations in the scene, such as cluttered grass/water plants and backgrounds with sharp intensity changes.
\texttt{Object Detection vs. Instance Segmentation:}
These two tasks show broadly similar allocation patterns, but with consistent differences.
For object detection, \method places relatively more compute on structural cues (reflected as stronger mid/high-frequency emphasis in FFT).
For instance segmentation, \method further increases attention to fine high-frequency details, such as sofa boundaries, sofa/wall/frame textures in the first example, and window edges and rail-like line structures in the second example.

\subsection{Efficiency Comparison}
We evaluate inference efficiency on the Kodak dataset using a single RTX 4090 with batch size 1.
We report total parameter count, activated parameter count, and end-to-end latency (encoder + decoder).
\method is not only more parameter-efficient, but also achieves the best practical coding latency, even outperforming the original TIC by 8.4\%.
\begin{table}[!t]
\centering
\caption{Efficiency comparison on kodak (TIC~\cite{lu2021transformerbasedimagecompression} as backbone.).}
\label{tab:efficiency}
\begin{tabular}{lccc}
\toprule
Method & Total Params (M) & Activated Params (M) & Latency (ms) \\
\midrule
TIC~\cite{lu2021transformerbasedimagecompression} & 7.51 & 7.51 & 155 \\
TransTIC~\cite{chen2023transtic} & 9.12 & 9.12 & 345 \\
AdapICMh~\cite{liu2023icmh} & 7.80 & 7.80 & 163 \\
\method & 7.17 & 6.25 & \textbf{142} \\
\bottomrule
\end{tabular}
\end{table}

\section{Conclusion}
we present \method, a unified multi-task image coding architecture, featured by it's sparsely activated and dynamic compute allocation character. \method provides a novel way for end-to-end multi-task codec training, against previous static and task-specific custom training paradigm. Extensive evaluation and ablation demonstrate the effectiveness of our proposed method.

%
%
\bibliographystyle{splncs04}
\bibliography{main}


\end{document}